\documentstyle{article}
\begin{document}
\newcommand{\nc}{\newcommand}
\nc{\be}{\begin{displaymath}}
\nc{\ee}{\end{displaymath}}
\nc{\ble}{\begin{equation}}
\nc{\ele}{\end{equation}}
\nc{\bea}{\begin{eqnarray*}}
\nc{\eea}{\end{eqnarray*}}
\nc{\bela}{\begin{eqnarray}}
\nc{\eela}{\end{eqnarray}}
\nc{\rbo}{\raisebox}
\nc{\cH}{{\cal H}}
\nc{\RR} {\rangle \! \rangle}
\nc{\LL} {\langle \! \langle}
\nc{\rmi}[1]{{\mbox{\small #1}}}
\nc{\eq}{eq.~}
\nc{\nr}[1]{(\ref{#1})}
\nc{\ul}{\underline}
\nc{\cM}{{\cal M}}
\nc{\mc}{\multicolumn}
\nc{\todo}[1]{\par\noindent{\bf $\rightarrow$ #1}}
\nc{\nonu}{\nonumber}
\nc{\lmax}{l_{\rm max}}
\nc{\half}{\mbox{\small$\frac12$}}
\nc{\eights}{\mbox{\small$\frac18$}}
\nc{\bg}{\bar \gamma}
\def\hga{{\widehat \gamma }}
\def\VVert{{|\! \! \; |\! \! \; | }}
\def\twtimes{{\widetilde \times }}
\def\lst{{l^*}}
\def\dst{{d^*}}
\def\bC{{\bf C}}
\def\bN{{\bf N}}
\def\bP{{\bf P}}
\def\bR{{\bf R}}
\def\bZ{{\bf Z}}
\def\cA{{\cal A}}
\def\cB{{\cal B}}
\def\cC{{\cal C}}
\def\cF{{\cal F}}
\def\cG{{\cal G}}
\def\cL{{\cal L}}
\def\cN{{\cal N}}
\def\cP{{\cal P}}
\def\cR{{\cal R}}
\def\cS{{\cal S}}
\def\cT{{\cal T}}
\def\cU{{\cal U}}
\def\cZ{{\cal Z}}
\def\om{{\overline m}}
\def\oU{{\overline U}}
\def\oX{{\overline X}}
\def\g{{\gamma}}
\def\L{{\Lambda}}
\def\rv{{\rm v}}
\def\eins{{\bf 1}}
\def\und#1{\underline{#1}}
\def\Gau#1#2{d\mu_{#1}({#2})}
\def\iGau#1#2{\int d\mu_{#1}({#2})}
\def\Gaug{d\mu_{\g}({\Phi })}
\def\Gaurv{d\mu_{\rv }({\Phi })}
\def\iGaug{\int d\mu_{\g}({\Phi })}
\def\GauN#1#2{d\mu_{#1}^{(N)}({#2})}
\def\iGauN#1#2{\int d\mu_{#1}^{(N)}({#2})}
\def\iGaugN{\int d\mu_{\g}^{(N)}({\Phi })}
\def\GaugN{d\mu_{\g}^{(N)}({\Phi })}
\def\GaurvN{d\mu_{\rv }^{(N)}({\Phi })}
\def\Ap{{A^{\prime }}}
\def\ap{{a^{\prime }}}
\def\Lp{{\Lambda^{\prime }}}
\def\Rp{{R^{\prime }}}
\def\Vp{{V^{\prime }}}
\def\Zp{{Z^{\prime }}}
\def\up{{u^{\prime }}}
\def\gp{{\g^{\prime }}}
\def\rhop{{\rho^{\prime }}}
\def\yp{{y^{\prime }}}
\def\inn{{\underline \in }}
%
\begin{titlepage}
\hfill MS-TPI-94-12 \\

\begin{centering}
\vfill

{\LARGE \bf On Renormalization\\
            Group Flows\\
            and\\
           \vspace{3mm}
            Polymer Algebras}
\renewcommand{\thefootnote}{\arabic{footnote}}

\vspace{2cm}
{\bf A. Pordt}\footnote{e-mail: pordt@yukawa.uni-muenster.de} \\[6mm]

\vspace{0.3cm}
{\em
\, Institut f\"ur Theoretische Physik I, Universit\"at M\"unster, \\
   Wilhelm-Klemm-Str.\ 9, D-48149 M\"unster, Germany \\}

\vspace{2cm}
{\bf Abstract} \\
\end{centering}
\vspace{0.2cm}
In this talk methods for a rigorous control of the renormalization
group (RG) flow of field theories are discussed.
The RG equations involve the flow of an infinite number of local
partition functions. By the method of exact beta-function the RG
equations are reduced to flow equations of a finite number of
coupling constants.
Generating functions of Greens functions are expressed by polymer activities.
Polymer activities are useful for solving the large volume and
large field problem in field theory.
The RG flow of the polymer
activities is studied by the introduction of polymer algebras.
The definition of products and recursive functions replaces
cluster expansion techniques.
Norms of these products and recursive functions
are basic tools and simplify a RG analysis for field theories.
The methods will be discussed at examples of the $\Phi^4$-model,
the $O(N)$ $\sigma$-model and hierarchical scalar field
theory (infrared fixed points).
\vfill \vfill
\noindent
MS-TPI-94-12\\
October 1994
\end{titlepage}
%
%
%
\section{Introduction}
There are several goals in constructive field theory. The first one is
a rigorous definition of Euclidean functional-integrals.
There exists many examples for a construction of special
functional-integrals (see e.~g.
\cite{GJ73,MS77,MP85,GJ87,FMRS87,GK83,GK85,B88a,B88b,BY90,DH91,R91,DH92,DH93}
).
But a general definition of the Euclidean
functional-integral is still lacking. The second goal in constructive field
theory is to compute
functional-integrals by approximation methods. The functional-integrals
can be represented by convergent series expansions in terms of
finite-dimensional integrals (cp.~\cite{MP89}). This is in analogy
to conventional
perturbation theory where functional-integrals are expressed
in terms of (finite-dimensional)
Feynman-integrals. The definition
and construction of functional-integrals lead to a proof of the existence of
ultraviolet- resp. infrared fixed points. The problem
is the infinite (not denumerable) number of degrees of freedom
which is connected to ultraviolet and infrared problems. The large
field problem is connected to the divergence of standard power series
expansions. A third aim is to represent the construction in such
a way that all intermediary steps can be done in a finite number
of well-defined computations.

There exists several tools to perform such a program. A field theory
which is represented by a functional-integral can be studied by means
of Wilson's renormalization group (RG) \cite{W71,WK74,W83}.
Thereby, the original functional-integral is
represented by a RG flow of effective functional-integrals. These
effective functional-integrals are simpler to define
than the original functional-integral. The RG flow
of the effective functional-integrals has to be controlled.
The effective functional-integrals can be further analyzed by
methods used in statistical mechanics, especially by the introduction of
polymer systems (cp. \cite{GrK71,MP85}). A suitable defined polymer
system can control large field contributions and solves the large
volume problem. Thereby, the effective systems represented by effective
functional-integrals will be decomposed into finite subsystems.
The effective systems
depend on an infinite number of degrees of freedom.
Their control is reduced to
the problem of analyzing the RG flow of finite subsystems.
In a
RG analysis, using conventional perturbation theory, one distinguishes
between
relevant and irrelevant parts. Likewise, there are relevant and
irrelevant parts of the effective finite subsystems.
The relevant part depends only on a finite
number of parameters. The flow of the irrelevant part can be controlled
by an application of fixed point theorems. Thereby, the control of the
RG flow
of effective systems is reduced to a RG flow defined in a finite-dimensional
parameter space. This method is called the {\em method of exact
beta-functions}. For a control of effective systems one introduces
norms for polymer activities.
A suitable definition of norms and polymer systems is the technical
core in the construction of field theoretic models.

It is the aim of this talk to review old and provide new
tools and definitions for
such a program.
A test for simplicity of these methods is the implementability on a
computer.

This paper is organized as follows. We start in Sect. 2 with
the introduction to general renormalization group transformations (RGT).
Then, we consider the special example of the Kadanoff-Wilson
(linear) block spin transformation. This RGT were firstly applied to field
theory by Gaw\c edzki and Kupiainen (cp.~\cite{GK80,GK83,GK85}).
Next, we consider the definition of the
nonlinear block spin RGT at the example of the nonlinear $O(N)$
$\sigma $-model.

Section 3 introduces a general polymer systems and presents the definition
of polymer activities by introducing an exponential function $EXP$.
The exponential function $EXP$ is defined by a product $\circ $. This
$\circ $-product was also used by Brydges and Yau, Dimock and Hurd
\cite{BY90,DH91} and has its origin in a
product defined for problems in statistical mechanics by Ruelle
\cite{Ru69}. Ruelle's product differs from the $\circ$-product used in
constructive
field theory by an important point\footnote{The author
thanks D.~Brydges for this comment}. It does not allow
an overlap of indices,
whereas the $\circ $-product does not allow an overlap of lattice points.
This property of nonoverlapping of lattice points is essential for a
control of large fields.
The RG flow of the effective systems is represented by the
RG flow of polymer activities. The effective
subsystems are defined on lattices.

One RG step can be decomposed into four steps.
The first step is called integration step. In this step high momentum
fields are integrated out. This can be done recursively,
using a $\times_\Gamma$-product.
After this step the correlation length becomes
larger. Then, the activities are defined on a coarser lattice
(coarsening step). The coarsening step can be performed in a recursive
way by introducing a mapping $E_A.$
The polymer acivities are defined in such a way that the localization
property holds. A polymer activity $A(P|\psi )$ for a polymer $P$
and field $\psi $ obeys the localization property if $A(P|\psi )$ depends
only on $\psi (y)$ for $y \in P$. This localization property of the
polymer activities makes a third step (localization step) necessary.
The localization step can be performed recursively, like the integration
step, by using a $\times_\cA$-product.
The fourth and last step rescales the fields such that the new effective
polymer activities lives on the same lattice as the polymer activities
before the RG step.

Section 4 presents the general method of exact beta-functions. As examples
we consider the RG flow of the $\Phi^4$-model and
hierarchical RG fixed points. Koch and Wittwer
\cite{KW86,KW91} applied
the method of exact beta-functions to construct the double-well
fixed point in 3 dimensions. These method is a candidate for the
construction of field theories with no small coupling constants.

The split into relevant and irrelevant parts and the method
of renormalization
and repolymerization for the flow of effective polymer systems are studied
in Sect. 5. For repolymerization a further RG step is necessary
(repolymerization step).

Section 6 presents a norm for the polymer activities and shows how
this norm behaves under RG steps. It will be shown that large
fields are controlled by the method of exponential pinning.
%
%
\section{Renormalization Group Transformations}
Our main object of interest is the generating functional of
Euclidean Greens functions (partition function).
It is the following infinite-dimensional integral
\be
           Z = \int [D\phi ]\, \cZ (\phi ), \qquad [D \phi ] =
               \prod_{z\in \bR^d} d\phi (z) \enspace ,
\ee
where $\cZ (\phi )$ is a real-valued function, called
{\em Boltzmannian}. We restrict our attention here
to real-valued scalar fields $\phi $. In renormalization group (RG)
investigations the computation of $Z$ is performed stepwise. Let us
consider the definition of one RG step. Define new fields $\Phi $
and a function $P(\Phi ,\phi )$ which obeys
\ble \label{Pdef}
           \int [D\Phi ]\, P(\Phi ,\phi ) =1 \enspace .
\ele
Then, the {\em renormalization group transformation} is defined as follows
\ble \label{RGT}
       \cZ^{\prime } (\Phi ) = \int [D\phi ]\, P(\Phi ,\phi ) \cZ (\phi )
        \enspace .
\ele
Eqs. (\ref{Pdef}) and (\ref{RGT}) imply
\be
           Z = \int [D\phi ]\, \cZ (\phi ) = \int [D\Phi ]\,
               \cZ^{\prime } (\Phi ) \enspace .
\ee
Thus the new Boltzmannian $\cZ^{\prime }$ and the new
field $\Phi $ can be used to compute the partition function $Z.$
This RG procedure
can be repeated and the result is the
{\em RG flow of effective Boltzmannians} :
\be
           \cZ \longrightarrow \cZ^{\prime} \longrightarrow
               \cZ^{\prime \prime } \longrightarrow \ldots \enspace .
\ee
The definition of the RG is chosen in such a way that the
effective Boltzmannians depend on fewer and fewer degrees of
freedom. Thus, the task of computing $Z$ by an infinite-dimensional
integral is solved by computing an infinite number of RG steps.
Instead of considering the RG flow of
effective Boltzmannians, it is better to consider the RG flow of effective
polymer activities. In this way the RG transformations can be represented by
finite-dimensional integrals.

Before coming to the definition of polymer activities, we will study two
examples of RG transformations. The first example is the {\em Kadanoff-Wilson
(linear) block spin transformation}. This method was applied
to field theory by Gaw\c edzki and Kupiainen \cite{GK80,GK83,GK85}.
The second example is the nonlinear
block spin transformation applied to the nonlinear $\sigma$-model.
For a definition of a RG transformation introduce new
fields $\Phi $ and the function $P(\Phi ,\phi )$ which obeys
eq. (\ref{Pdef}). Let us suppose that the original field $\phi $ lives
on the lattice $\L := (a\bZ )^d$ and define the integral
for the partition function $Z$ by  $[D\phi ] := \prod_{y\in \L} d\phi (y)$.
Decompose the lattice $\L $ into hypercubes of side length $La$,
where $L$ is a fixed number, $L\in \{ 2,3,\ldots \} .$
The center points of these hypercubes
(blocks) build also a lattice with side length $La.$
This lattice is called {\em block lattice} $\Lp := (La\bZ )^d.$  A site
$y$ of the lattice $\L $ is contained in a site $x$ of the block lattice
$\Lp $ if $y$ is contained in the block with center point $x.$ In this case
we write $y\inn x.$ For a field $\phi : \L \rightarrow \bR ,$
define the {\em block spin field} : $C\phi : \Lp \rightarrow \bR $ by
\be
           C\phi (x) := \beta \sum_{y:\, y\inn x} \phi (y), \qquad
                         x\in \Lp \enspace .
\ee
$\beta $ is a positive real number, called {\em scaling parameter}.
Finally, the special RG transformation, called {\em block spin
transformation}, is defined by
\be
           \cZ^{\prime } (\Phi ) = \int [D\phi ]\,
             P_{\alpha ,\beta }(\Phi ,\phi ) \cZ (\phi )\enspace ,
\ee
where $P_{\alpha ,\beta }$ is defined by
\be
           P_{\alpha ,\beta }(\Phi ,\phi ) :=
            \cN_{\alpha ,\beta }\, \exp \{ -\frac{\alpha}{2}
               \sum_{x\in \Lp} (\Phi (x) -(C\phi )(x))^2\} \enspace .
\ee
$\alpha $ is a positive real parameter and $\cN_{\alpha ,\beta }$
is a normalization constant such that eq. (\ref{Pdef}) holds.

We represent the Boltzmannian by a free propagator $u$ and interaction $V$
\be
           \cZ (\phi ) = \exp \{ -\frac{1}{2}(\phi ,u^{-1}\phi ) -
                           V(\phi ) \} \enspace .
\ee
$(\cdot ,\cdot )$ is the canonical bilinear form. Then, the RG transformation
reads
\be
           \cZ^{\prime } (\Phi ) = \cN_{\alpha ,\beta }\,
             \exp \{ -\frac{1}{2}(\Phi ,\up^{-1}\Phi )\}
           \, \exp \{ -\Vp (\Phi ) \} \enspace ,
\ee
where the {\em effective interaction} $\Vp $ is defined by
\be
             \exp \{ -\Vp (\Phi ) \} :=
               \int [D\zeta ]\, \exp \{ -\frac{1}{2}(\zeta ,
                 \Gamma^{-1}\zeta )\}
                   \exp \{ -V(\zeta + \cA \Phi ) \} \enspace .
\ee
The {\em fluctuation propagator} $\Gamma $ and the {\em
block spin propagator} $\up $ are
\be
\Gamma := (u^{-1} + \alpha C^TC)^{-1},
\qquad
\up := (\alpha - \alpha^2C\Gamma C^T)^{-1}\enspace .
\ee
The $\cA $-operator, which maps a field defined on $\Lp $ to fields
defined on $\L ,$ is given by $\cA := uC^T\up^{-1}.$
The normalized Gaussian measure with mean zero is defined by
\be
d\mu_\Gamma (\zeta ) := \cN \, [D\zeta ]\,
               \exp \{ -\frac{1}{2}(\zeta ,\Gamma^{-1}\zeta )\} \enspace .
\ee
The RG transformation, for the effective interaction $V,$ reads
\be
           e^{-\Vp (\Psi ) } = \iGau{\Gamma}{\phi} e^{-V(\phi + \cA
                \Psi )} \enspace .
\ee
Let the free propagator $u$ be a Gaussian fixed point. Then self-similarity
holds
\be
           \up (Ly, L\yp ) = L^{2-d(+\epsilon )} u(y,\yp ), \qquad
              y,\yp \in \L \enspace .
\ee
Replacing the field $\Phi (\cdot )$ by the rescaled one,
$L^{1-\frac{d}{2}(+\frac{\epsilon}{2})} \Phi (\frac{\cdot}{L}),$ we obtain
the RG transformation after rescaling
\be
           e^{-\Vp (\Phi ) } = \iGau{\Gamma}{\zeta} e^{-V(\zeta +
             L^{1-\frac{d}{2}(+\frac{\epsilon}{2})}\cA
                (\Phi (\frac{\cdot}{L})))}\enspace .
\ee
The field $\Phi $ lives on the original lattice $\L $ but with reduced
correlation length. This procedure can be iterated and the result is the
following RG flow
\be
e^{-V} \longrightarrow  e^{-\Vp} \longrightarrow
                  e^{-V^{\prime \prime}} \longrightarrow \cdots \enspace .
\ee
Best localization properties are obtained for $\alpha = O(1)$
(see Bell and Wilson \cite{BW74}).

We consider a second example for a RG transformation.
This is the nonlinear block spin transformation at the example of the
nonlinear $O(N)$ $\sigma $-model. Define a measure on the
$N-1$-dimensional unit sphere $S^{N-1}$
\be
        \int_{S^{N-1}}[d\sigma ] := \int \prod_{z\in \L }
        d\sigma_1(z) \cdots d\sigma_N(z) \delta (\sigma^2(z)-1)
\ee
and a partition function
\be
Z = \int_{S^{N-1}}[d\sigma ] e^{-V(\sigma )} \enspace ,
\ee
where the interaction $V$ is defined by nearest-neighbor couplings
\be
        V(\sigma ) := \beta \sum_{\mu =1}^d \sum_{y\in \L } \left(1-
        \sigma (y) \cdot \sigma (y+\hat{\mu } )\right) \enspace .
\ee
$\hat{\mu } $ is a vector in $\mu $-direction with length of one
lattice spacing $a.$
Let $\mu :\Lp \rightarrow S^{N-1} $ be a unit vector field on the block
lattice $\Lp .$ Then, the nonlinear block spin RG transformation
is defined by
\be
        \cZ^{\prime} (\mu ) := \int_{S^{N-1}} [d\sigma ] P(\mu ,\sigma )
                             e^{-V(\sigma )}\enspace ,
\ee
where
\be
        P(\mu ,\sigma ) := \exp \{ -W_\kappa (\sigma )\} \,
          \prod_{x\in \Lp } \exp \{ \beta \kappa \mu (x) \cdot
            \sum_{y\inn x} \sigma (y) \} \enspace .
\ee
$W_\kappa $ is defined such that
\be
        \int_{S^{N-1}}[d\mu ] P(\mu ,\sigma ) = 1
\ee
holds. A simple computation shows
\bea
        W_\kappa (\sigma ) &=& \sum_{x\in \L } \ln \left(
         \int_{S^{N-1}} [d\mu ]\, \exp \{ \kappa \beta \mu \cdot
           \sum_{y\inn x} \sigma (y) \} \right) \nonu\\
          &=& const + \sum_{x\in \L } \ln \left(
            \frac{ I_{\frac{N-2}{2}}(\kappa \beta
              |\sum_{y\inn x} \sigma (y)| )}{(\kappa \beta |
                \sum_{y\inn x} \sigma (y)|)^{\frac{N-2}{2}}}\right) \enspace .
\eea
$I_\nu (z)$ is the Bessel function represented by the series expansion
\be
        I_\nu (z) = \frac{z^\nu}{2^\nu}\, \sum_{k=0}^\infty (-1)^k
           \frac{z^{2k}}{2^{2k}k!\Gamma (\nu + k + 1)} \enspace .
\ee

We have seen how to compute partition functions iteratively
by the introduction
of RG transformations. RG transformations are given by infinite-dimensional
integrals. For a computation of RG steps one has to reduce these
infinite-dimensional integrals to finite-dimensional ones. This problem
corresponds
to the infinite volume problem in statistical mechanics and can be
solved by cluster expansion methods (cp.~\cite{B84}) or equivalently
by the introduction of polymer systems.
%
%
\section{Polymer Systems, Activities, Exponentiation}
Polymer systems for statistical mechanics were introduced by
Gruber and Kunz \cite{GrK71}. This section presents a polymer system
for the use of a RG analysis for field-theoretic models.
The definition of polymer activities presented here uses block spin
RG transformations. The difference to other definitions of RG
transformations is the introduction of the $\cA $-operator. An
advantage in using block spin RG is that new polymer activities are
defined by finite-dimensional integrals. A further advantage is
that gradients of fields can be represented by gradients of the
$\cA $-operator, $\partial (\cA \psi ) = (\partial \cA )\psi .$
For estimations of such gradients one has to bound gradients of $\cA $
and there is no need of Sobolev-inequalities.
A disadvantage is that the $\cA $-operator is non-local and has to
be taken into consideration for a RG step of polymer activities.
A definition, not using block spin RG,
similar to the one presented here can be found in
Brydges contribution to these proceedings and in
Brydges and Yau \cite{BY90} and Dimock and Hurd \cite{DH91,DH92,DH93}.

Let $\L = (a\bZ )^d$ be the $d$-dimensional hypercubic lattice with
lattice spacing $a.$ For a subset $P$ of $\L $ denote by $|P|$
the number of elements in $P.$ The union $P\cup Q$ is denoted by
$P+Q$ if $P$ and $Q$ are disjoint sets. The set consisting of all
elements which are in $P$ but not in $Q$ is denoted by $P-Q.$ Let
$Pol(\L )$ be a subset of the set of all finite subsets of $\L $,
$\cP_{fin}(\L ) := \{ P\subseteq \L |\, |P|<\infty \} ,$ such that
$P,Q\in Pol(\L )$ implies $P+Q, P-Q\in Pol(\L ).$
$Pol(\L )$ is called a {\em set of polymers.}
Let $\Lp $ be a block lattice of $\L .$ Suppose we have also defined
a set of polymers $Pol(\Lp )$. For simplicity suppose here that
$Pol(\L ):=\cP_{fin}(\L )$ and $Pol(\Lp ):=\cP_{fin}(\Lp )$, i.e.
polymers are finite subsets.
We want to truncate the fluctuation propagator $\Gamma $
and the $\cA $-operator on polymers. For a polymer $P$ let
$\chi_P $ be the characteristic function. For $P\in Pol(\L )$ and
$X\in Pol(\Lp )$ define truncated operators
\be
         \Gamma_P:= \chi_P \Gamma \chi_P,\qquad  \cA_X :=\cA \chi_X\enspace .
\ee
For a polymer $X\in Pol(\Lp )$ let us define a polymer $\oX \in Pol(\L ) $
by
\be
      \oX := \{ y\in \L |\, \exists x\in X:\, y\inn x\}
\ee
and for a polymer $Y\in Pol(\L )$ a polymer $[Y] \in Pol(\Lp ) $ by
\be
      [Y] := \{ x\in \Lp |\, \exists y\in Y:\, y\inn x\} \enspace .
\ee
Consider the {\em set of effective polymer partition functions} :
\ble \label{DefPPF}
        \cZ (\L ,\cF ) := \{ Z:\, Pol(\L )\times \cF \rightarrow \bR |\,
                             Z(\emptyset |\Psi ) = 1 \} \enspace ,
\ele
where
$\cF $ is the set of fields, e.g. $\cF := Fun(\L ):= \{ F:\L \rightarrow
\bR \} .$
The RG transformation (without rescaling)
for effective polymer partition functions is defined by
\be
  \Zp (X|\Psi ) := \iGau{\Gamma_\oX}{\zeta } Z(\oX |\zeta +\cA_X\Psi )
   \enspace ,
\ee
where $Z\in \cZ (\L ,Fun(\L ))$ and $\Zp \in \cZ (\Lp ,Fun(\Lp )).$
The RG transformation with rescaling is a mapping
$\cR \cG :\, \cZ (\L ,Fun(\L ))
\rightarrow \cZ (\L ,Fun(\L ))$, $\Zp = \cR \cG (Z).$

The RG transformation can be performed in 4 steps. The first step
is the {\em integration step} defined by the
mapping
\be\mu_\Gamma :\, \cZ (\L ,Fun(\L )) \rightarrow
                                          \cZ (\L ,Fun(\L )),\quad
        \mu_\Gamma (Z)(Y|\phi ) := \iGau{\Gamma_Y}{\zeta} Z(Y|\zeta +\phi )
        \enspace .
\ee
The second step is the {\em coarsening step} defined by the mapping
\be
[\ ] :\, \cZ (\L ,Fun(\L ))\rightarrow
              \cZ (\Lp ,Fun(\L )),\quad
        [Z](X|\phi ) := Z(\oX |\phi )\enspace .
\ee
The third step is the {\em localization step} defined by the mapping
\be
\iota_\cA :\, \cZ (\Lp ,Fun(\L ))\rightarrow
                 \cZ (\Lp ,Fun(\Lp )), \quad
        \iota_\cA (Z)(X|\Psi ) := Z(X|\cA_X \Psi )\enspace .
\ee
The fourth and last step is the {\em rescaling step} defined by the mapping
\be\cR_L:\, \cZ (\Lp ,Fun(\Lp ))\rightarrow
                 \cZ (\L ,Fun(\L )), \quad
        \cR_L (Z)(Y|\phi ) := Z(LY|L^{1-\frac{d}{2}}\phi (\frac{\cdot}{L}))
          \enspace .
\ee
The mapping $\cR \cG $ is therefore
a composition of the four above defined mappings,
$\cR \cG := \cR_L \circ \iota_\cA \circ [\ ] \circ \mu_\Gamma $
\bea
   \cZ (\L ,Fun(\L )) &\stackrel{\mu_\Gamma}{\rightarrow}&
     \cZ (\L ,Fun(\L ))\stackrel{[\ ]}{\rightarrow}
       \cZ (\Lp ,Fun(\L ))
\nonu\\
       &\stackrel{\iota_\cA}{\rightarrow}&
         \cZ (\Lp ,Fun(\Lp )) \stackrel{\cR_L}{\rightarrow}
           \cZ (\L ,Fun(\L ))\enspace .
\eea
Polymer partition functions obey the following conditions :
\par\vspace{0.25cm}
\begin{itemize}
\item[(a)] Locality :
           \be
           \frac{\partial }{\partial \Psi (y)}
           Z(P|\Psi ) =0,\quad \forall y \notin P
           \ee
\item[(b)] Euclidean lattice symmetry :
           \be
           Z(RY|R\Psi ) =
           Z(Y|\Psi ),
           \ee
           $\quad \forall R\in $ group of lattice symmetry
\item[(c)] Invariance under external symmetry transformations\\
           $\Psi \rightarrow U\Psi $ :
           \be
           Z(Y|U\Psi ) = Z(Y|\Psi )
           \ee
\item[(d)] Approximation (thermodynamic limit):
           \be
           \lim_{Y\nearrow \L } Z(Y|\Psi ) =
           Z (\Psi )
           \ee
\end{itemize}
For a control of polymer partition functions it is better to
change to polymer activities. Polymer activities depend, like the polymer
partition functions, on polymers and fields. The set of polymer
activities is equivalent to the set of polymer partition functions.
This means that
if we know the polymer activities, then the polymer partition
functions are determined and vice versa. The polymer activities have
the following important property (at least for weakly coupled models)
which the polymer partition functions do not share. The value of a polymer
activity for a given polymer and field is small if the polymer
contains a large
number of elements or is of large extension. The polymer activities
and polymer partition functions are related by an exponential function
$EXP.$ To define this function we will introduce a product $\circ $.
The space, where this product is defined, is the {\em
set of all polymer functions}
defined by
\be
        \cM (\L ,\cF ) :=
        \{ F:\, Pol(\L )\times \cF \rightarrow \bR \} \enspace ,
\ee
where $\cF $ is a set of fields.
For notational simplicity we omit below the field-dependence of
the polymer function. Define the $\circ$-product,
for two polymer functions $U, V\in \cM ,$ by
\ble \label{CirPro}
        (U\circ V)(Y) := \sum_{P_1,P_2\in Pol(\L ):\atop P_1+P_2=Y}
                           U(P_1) V(P_2)\enspace .
\ele
The sum is over all partitions of $Y$ into two disjoint subsets.
Defining an addition $+$ on $\cM $ in the canonical way, we see that
$(\cM ,\circ ,+)$ is an associative algebra with unit element $\eins $,
$\eins (P) := \delta_{P,\emptyset}$. Call this algebra a {\em
polymer algebra }.
Define on $\cM $ a $\cdot $-multiplication by
\ble \label{DefPro}
(U\cdot V)(Y) := U(Y) V(Y)\enspace .
\ele
The set of all polymer activities $\cA $ is a subset of the set of
polymer functions $\cM ,$ defined by
\be
        \cA (\L ,\cF ) := \{ A\in \cM |\,
        A(\emptyset |\Psi ) = 0\} \enspace .
\ee
$\cZ , \cA \subset \cM $ are subalgebras of $(\cM ,\circ ,+).$

We are now in the position to define the exponential mapping
$EXP:\, \cA \rightarrow \cZ $, where $\cZ $ is the set of all
polymer partition functions defined by eq.~(\ref{DefPPF}).
For a polymer activity $A\in \cA ,$ define
\ble \label{DefExp}
        EXP(A) := \eins + \sum_{n:\, n\ge 1} \frac{1}{n!}
           \underbrace{A\circ \cdots \circ A}_{n-times}\enspace .
\ele
$EXP$ is bijective with inverse mapping $LN := EXP^{-1}$. Furthermore,
$EXP:\, (\cA ,+) \rightarrow (\cZ ,\circ )$ is a group-isomorphism, i.~e.
\be
EXP(A+B) = EXP(A) \circ EXP(B)\enspace .
\ee
Let $Y\in Pol(\L )-\{ \emptyset \}$ be a nonempty polymer. Define
the {\em set of all partitions} of $Y$ by
\bea
        \Pi (Y) &:=& \bigcup_{n:\, n\ge 1} \Pi_n(Y),\nonu\\
          \Pi_n(Y) &:=& \left\{ \{ P_1,\ldots ,P_n\}
          |\, Y=P_1+\cdots +P_n\right\} \enspace .
\eea
The definition (\ref{DefExp}) of the exponential-function $EXP$
and the fact that $(\cM ,\circ ,+)$ is an associative algebra imply
the following explicit representations
\bea
        EXP(A)(Y) &=& \sum_{\bP \in \Pi (Y) } \prod_{P\in \bP } A(P)
      \nonu\\
        LN(Z)(Y)  &=& \sum_{n:\, n\ge 1} (-1)^{n-1} (n-1)!
          \sum_{\bP \in \Pi_n (Y) } \prod_{P\in \bP } Z(P)\enspace .
\eea
For the proof that $EXP$ is a bijective mapping one uses
\be
A(Y) = EXP(A)(Y) - \sum_{\bP \in \Pi (Y) \atop P\in \bP :\, |P|<|Y|}
\prod_{P\in \bP } A(P)
\ee
and proceeds by induction in the number of elements of a polymer.
The RG transformation $\cR \cG $ for polymer partition function
can be transformed to a RG transformation $\widetilde{\cR \cG}$
for polymer activities
\be
        \widetilde{\cR \cG }:= EXP^{-1} \circ \cR \cG \circ EXP :\,
          \cA \rightarrow \cA \enspace .
\ee
This shows that $\cR \cG $ and $\widetilde{\cR \cG }$ are equivalent
RG transformations.
The RG flow of partition functions and activities are related
by the following commutative diagram
\be
        \begin{array}{*{10}c}
          \  &
          \cZ & \stackrel{\hbox{\normalsize $\mu_\Gamma$}}{\rightarrow} & \cZ &
            \stackrel{\hbox{\normalsize $[\ ]$}}{\rightarrow} & \cZ &
            \stackrel{\hbox{\normalsize $\iota_\cA$}}{\rightarrow} & \cZ
          & \stackrel{\hbox{\normalsize $\cR_L$}}{\rightarrow} & \cZ \\
          {\hbox{\small $EXP$}} &
          \uparrow & {\hbox{\small $EXP$}} & \uparrow & {\hbox{\small $EXP$}}
          & \uparrow & {\hbox{\small $EXP$}}  & \uparrow &
          {\hbox{\small $EXP$}} & \uparrow \\
          \ &
          \cA & \stackrel{\hbox{\normalsize
$\widetilde{\mu_\Gamma}$}}{\rightarrow}
          & \cA & \stackrel{\hbox{\normalsize $\widetilde{[\ ]}$}}{\rightarrow}
          & \cA & \stackrel{\hbox{\normalsize
$\widetilde{\iota_\cA}$}}{\rightarrow}
          & \cA & \stackrel{\hbox{\normalsize
$\widetilde{\cR_L}$}}{\rightarrow} & \cA
        \end{array}
\ee
For a control of the RG flow of the polymer activities one has to study the
four mappings $\widetilde{\mu_\Gamma}$, $\widetilde{[\ ]}$,
$\widetilde{\iota_\cA}$ and $\widetilde{\cR_L}$ in more detail.

\subsection{Integration Step}
New polymer activities $\Ap = \widetilde{\mu_\Gamma}(A)$ can be
explicitly represented by tree graph formulas (cp.~\cite{B84,P86}).
Instead of presenting here tree graph formulas, we introduce a recursive
procedure by definition of a product (cp.~ also \cite{BY90}).

For a representation of the mapping $\widetilde{\mu_\Gamma}$
one introduces the set of pa\-ra\-me\-tri\-zed polymer activities
\be
        \cA (\L , [0,1], \cF ) := \{ U :\, Pol (\L ) \times [0,1] \times \cF
        \rightarrow \bC \}
\ee
and special parametrized polymer activities $A^{\prime}_s$, $s\in [0,1]$, by
\ble \label{ParDef}
        EXP(A^{\prime}_s) = \mu_{s\Gamma} (EXP(A))\enspace .
\ele
This implies
\be
        A^{\prime}_0 = A,\qquad A^{\prime}_1 = \widetilde{\mu_\Gamma}(A)
          \enspace .
\ee
Define on the set of parametrized polymer activities a
$\times_\Gamma$-product. For two parametrized polymer activities
$A^{\prime}_s, B^{\prime}_s\in
\cA (\L ,[0,1],\cF )$ define the product
\bela \label{timDef}
        \lefteqn{
        (A^{\prime} \times_\Gamma B^{\prime })_t(Y|\Psi ) :=}
        \nonu\\ & &
        \frac{1}{2} \sum_{Y_1,Y_2:\atop Y_1+Y_2 = Y} \sum_{y_1\in Y_1
        \atop y_2 \in Y_2} \int_0^t ds\, \iGau{(t-s)\Gamma}{\phi}
        \nonu\\ & &
        \frac{\partial}{\partial \Psi (y_1)} A^{\prime}_s(Y_1|\phi +\Psi )
        \Gamma (y_1, y_2)
        \frac{\partial}{\partial \Psi (y_2)} B^{\prime}_s(Y_2|\phi +\Psi )
         \enspace .
\eela
We want to compute the parametrized activity $A^{\prime}_t(Y|\Psi )$ by
recursion in the number of elements of $Y.$
Use the following definitions
\be
        \eins_n(Y) := \left\{ \begin{array}{r@{\quad:\quad}l}
                        1 & |Y| =n, \\ 0 & |Y| \ne n
                              \end{array} \right.
        \qquad
        \eins_{>n} := \sum_{k:\, k>n} \eins_k \enspace .
\ee
Then, the recursive equation is
\ble \label{IntRec}
        A^{\prime}_t = \eins_{>1}\cdot \left[
          (A^{\prime} \times_\Gamma A^{\prime })_t + A \right] +
           \eins_1 \cdot \mu_{t\Gamma}(A)\enspace .
\ele
The $\cdot $-product is defined by eq.~(\ref{DefPro}).
The proof of eq.~(\ref{IntRec}) can be done in the following way.
Firstly, distinguish the parametrized polymer activities
$A^{\prime}_t,$ $t\in [0,1]$ defined by
eq.~(\ref{IntRec}) and eq.~(\ref{ParDef}). Then show that they are equal.
For $t=0$, eq.~(\ref{IntRec}) obeys $A^{\prime}_0 = A.$ By differentiation
of eq.~(\ref{ParDef}) with respect to $s$, we derive a first order
differential equation of $A^{\prime}_s$. Differentiation of eq.~(\ref{IntRec})
with respect to $t$ implies that $A^{\prime}_t$, defined by
eq.~(\ref{IntRec}), obeys the same differential equation. The initial
values at $t=0$ are the same. Therefore, the parametrized
polymer activities defined
by eq.~(\ref{ParDef}) and (\ref{IntRec}) are the same. Thus,
eq.~(\ref{IntRec}) is proven for all $t\in [0,1].$

In the definition of the $\times_\Gamma$-product eq.~(\ref{timDef}) the
right hand side depends only on polymers which contain lesser elements
than the polymer on the left hand side of eq.~(\ref{timDef}). This implies
that applying the activities in eq.~(\ref{IntRec}) to special polymers,
the recursion can be solved by a finite number of steps.

\subsection{Coarsening Step}
For polymer activities $A$ on the lattice $\L ,$ the {\em coarsened
polymer activities} $\widetilde{[A]}$ are defined on the
block lattice $\Lp $ by
\be
        EXP(\widetilde{[A]}) = [EXP(A)]\enspace .
\ee
For a recursive computation of the coarsened polymer activities
$\widetilde{[A]}$ define the following mapping $E_A :\cA
\rightarrow \cA $, $A\in \cA $
\bea
      E_A(B)(Y) &:=& A(Y) + \sum_{P:\, y\in P\subseteq Y}
        \sum_{n:\, n\ge 1} \frac{1}{n!} \nonu\\
         & & \sum_{P_1,\ldots ,P_n:\, [P]\cap [P_a] \ne \emptyset \atop
                    P+P_1+\cdots +P_n = Y \ [P_a]\cap [P_b] =
                      \emptyset ,\, a\ne b}
         A(P)\prod_{i=1}^n B(P_i) \enspace .
\eea
In the definition of the mapping $E_A$ we have chosen
an element $y\in Y,$ for all polymers $Y.$ This choice of $y$ is arbitrary.
Define the polymer activity $\cC A \in \cA (\L ,\cF )$ by the following
recursive equation
\ble \label{CoaRec}
        \cC A = E_A (\cC A)\enspace .
\ele
That this equation defines $\cC \cA $ recursively can be shown in
the following way. Apply the activities in eq.~(\ref{CoaRec}) to
a polymer $Y.$ Then, $\cC \cA (Y)$ can be expressed by a sum of products
of terms $\cC \cA (P),$ where $P$ are polymers with $|P| < |Y|.$
Thus, if $\cC \cA (P)$ is defined for all polymers $P$ containing
less than $N=|Y|$ elements, then $(\cC \cA )(Y)$ is defined by
eq.~(\ref{CoaRec}).
The recursion starts with the monomer (=polymer with only one element)
$\cC A (\{ y\} ) := A(\{ y\} )$ , $y\in\L .$ Then, the coarsened polymer
activity $\widetilde{[A]}(X)\in \cA (\Lp ,\cF )$ is
\be
\widetilde{[A]}(X) = (\cC A)(\oX )\enspace .
\ee
This equation can be proven in the following way. Firstly, it can be shown
that the coarsened polymer activity obeys
\be
\widetilde{[A]}(X) = \sum_{\bP \in \Pi (\oX ): \atop \gamma
(\{[P]|P\in \bP \} )\  connected} \prod_{P\in \bP } A(P)\enspace ,
\ee
where, for $P_1, \ldots ,P_n$, the Venn-diagram $\gamma (P_1, \ldots ,P_n)$
is defined in the following way. Each polymer $P_a$ is
represented by a vertex
and draw a line $(P_aP_b)$, $a\ne b$ if $P_a$ and $P_b$ are not disjoint,
$P_a\cap P_b \ne \emptyset .$ Take an arbitrary element $y$ of $\oX $
and take the polymer $Y\in \bP $ which contains the element $y.$
Then, the Venn-diagram $\gamma (\{[P]|P\in \bP \} - \{ Y\}  )$
decomposes into
connected components. This decomposition implies the recursive
equation (\ref{CoaRec}).

\subsection{Localization Step}
The localization mapping $\widetilde{\iota_\cA } :\, \cA \rightarrow
\cA $ obeys, for all polymers $X\in Pol(\Lp ),$
\be
        \sum_{X=\sum Q} \prod_Q A(Q|\cA_X \phi ) =
         \sum_{X=\sum Q} \prod_Q \widetilde{\iota_\cA} (A)(Q|\phi )\enspace .
\ee
Define a parametrized polymer activity $A^{\prime}_s (Q|\phi ,\psi )$,
for all $Q\in Pol(\Lp )$, and two fields $\psi $ and $\phi ,$ implicitly by
\be
        \sum_{X=\sum Q} \prod_Q A(Q|s(\cA_X \phi -\psi ) + \psi ) =
         \sum_{X=\sum Q} \prod_Q A^{\prime}_s(Q|\phi ,\psi )\enspace .
\ee
The parametrized polymer activity $A^{\prime}_s$ obeys, for the
special values $s=0,1,$
\be
      A^{\prime}_0(Q|\phi ,\psi ) = A(Q|\psi ), \qquad
      A^{\prime}_1(Q|\phi ,\psi ) = \widetilde{\iota_\cA} (A)(Q|\phi )
        \enspace .
\ee
For two parametrized polymer activities $A^{\prime}_s$ and
$B^{\prime}_s$ define the $\times_\cA$-product by
\bea
        \lefteqn{
        (A^{\prime} \times_\cA B^{\prime })_t(X|\phi ,\psi ) :=}
        \nonu\\ & &
        \frac{1}{2} \sum_{X_1,X_2:\atop X_1+X_2 = X}
        \sum_{y\in \overline{X_1}
        \atop x \in X_2} \int_0^t ds\,
        \cA (y,x) \phi (x)
        \frac{\partial}{\partial \psi (y)}
        \nonu\\ & &
        A^{\prime}_s\left(X_1|\phi ,
        (t-s)(\cA_X \phi - \psi )+\psi \right)
        \nonu\\ & &
        B^{\prime}_s\left(X_2|\phi , (t-s)(\cA_X \phi - \psi )+\psi \right)
        \enspace .
\eea
Then, the following recursive equation holds
\be
        A^{\prime}_t = \eins_{>1}\cdot \left[
          (A^{\prime} \times_\cA A^{\prime })_t + A \right] +
           \eins_1 \cdot \iota_{\cA ,t}(A)\enspace .
\ee
We have seen how to control the RG steps given by the mappings
$\widetilde{\mu_\Gamma},$ $\widetilde{[\ ]}$ and $\widetilde{\iota_\cA}$
for the RG flow of polymer activities $A$ by introducing products
$\times_\Gamma ,$ $\times_\cA $ and a recursive mapping $E_{A}.$
A control of the RG flow of polymer activities $A$ is achieved by
splitting $A$ into a relevant and an irrelevant part. This method
of exact beta-function is discussed in the next section.
%
%
\section{Exact Beta-Function Method}
In general, a RG transformation $\Zp = \cR \cG (Z)$ maps a partition
function $Z,$ which is described by an infinite number of parameters
to a partition function $\Zp ,$ which depends also on a infinite number
of degrees of freedom. The exact beta-function method
reduces the
RG flow of an infinite number of parameters to a finite number of
parameters called {\em coupling constants}
(see \cite{P93} for an applicication to hierarchical models).
The remaining infinite number of
parameters which determine the partition functions can be controlled
by the coupling constants. The extraction of a finite number of
coupling constants is done by a projection operator ${\rm P}$ such that
\be
        Z^{rel} = Z^{rel}(\g_0,\ldots ,\g_N) = {\rm P}(Z)
\ee
depends on a finite number of parameters $\g_0,\ldots ,\g_N.$
The {\em irrelevant part} of the partition function $Z$ is
\be
        R=(\eins -{\rm P})(Z)\enspace .
\ee
The RG flow of this irrelevant part $R$ can be controlled by
standard fixed point theorems. The irrelevant fixed point $R^* =
R^*(\g_0,\ldots ,\g_N)$ depends on the coupling constants
$\g_0,\ldots ,\g_N$ and is defined by
\be
        R^* = H_{\g_0,\ldots ,\g_N}(R^*) := (\eins -{\rm P})\cR \cG (Z^{rel}(
\g_0,\ldots ,\g_N) + R^*) \enspace .
\ee
Then, the exact beta-function $B:\, \bR^{N+1} \rightarrow \bR^{N+1}$ is
defined by
\bea
        \lefteqn{Z^{rel} (B(\g_0,\ldots ,\g_N)) =} \\
          & & {\rm P} \cR \cG \left(
           Z^{rel}(\g_0,\ldots ,\g_N) +
           R^* (\g_0,\ldots ,\g_N)\right) \enspace .
\eea
The problem of searching fixed points $Z^* =\cR \cG (Z^*)$ of
the RG transformation is solved by finding fixed
points $\g_0^*,\ldots ,\g_N^*$ for the exact beta-function $B$,
$B(\g_0^*,\ldots ,\g_N^*) =(\g_0^*,\ldots ,\g_N^*).$ Then,
the fixed point $Z^*$ is the sum of the relevant and irrelevant
part at $(\g_0^*,\ldots ,\g_N^*)$
\be
        Z^* = Z^{rel}(\g_0^*,\ldots ,\g_N^*) +
        R^* (\g_0^*,\ldots ,\g_N^*)\enspace .
\ee
For a control of the RG flow it is not necessary to determine the
irrelevant fixed point $R^*$ exactly. It is sufficient to
find a neighborhood $U=U(R^*)$ of $R^*$
which is stable under RG transformation,
$H_{\g_0,\ldots ,\g_N} (U) \subseteq U.$ The beta-function method
reduces the infinite-dimensional RG flow to a finite-dimensional
RG flow of running coupling constants $\g =(\g_0,\ldots ,\g_N)$ :
\be
        \g \stackrel{B}{\rightarrow}\g^{\prime} \stackrel{B}{\rightarrow}
          \g^{\prime \prime} \stackrel{B}{\rightarrow} \cdots \enspace .
\ee
In the follwing subsections we will discuss the RG flow at the examples
of the $\Phi^4$-model and hierarchical fixed points.
\subsection{Example: $\Phi^4_d$-Model}
Consider the RG transformations
\be
V_{j-1}(\Psi ) := -\ln \iGau{v^j}{\Phi } \exp \{ -V_j(\Phi +\Psi )\} -
                  (\Psi =0) \enspace ,
\ee
for all $j\in \{ 0,\ldots ,n\} ,$ $n\in \bN .$ The effective interactions
$V_j$ are normalized by subtraction of a constant term, such that $V_j(0)=0.$
$V_j$ is the effective interaction after $n-j$ RG steps. $v^j$ is the
fluctuation propagator for the $(n-j+1)$th RG step.
The starting (bare) interaction for the $\Phi^4$-model in $d$ dimensions
is given by
\be
V_n(\Phi ) = \frac{1}{2} m_n^2 a_n^{-2} \int_z \Phi^2(z) +
                \frac{1}{2} \beta_n \int_z (\nabla \Phi )^2(z) +
                  \frac{1}{4!} \lambda_n a_n^{d-4} \int_z \Phi^4(z)
                   \enspace ,
\ee
where $a_n:= L^{-n}a$, $a=$ unit length, $L\in \{ 2,3,\ldots \} .$
$m_0^2$, $\beta_n,$ and $\lambda_n$ are called bare coupling constants.
Since the field $\Phi $ has dimension
$a^{1-\frac{d}{2}}$, $\nabla $ has dimension $a^{-1}$ and $\int_z$ is of
dimension $a^d,$ we see that $\int_z \nabla^m\Phi^{2n}$ is of dimension
$a^{2n-m-(n-1)d}.$ Therefore, the constants $m_n,\beta_n$ and $\lambda_n$
are dimensionsless. Running coupling constants $m_j^2$, $\beta_j$,
$\lambda_j$ and $r_j$, $j\in \{ 0,\ldots ,n\} $ are defined by the following
representation of the effective interactions
\bea
V_j(\Phi ) &=& \frac{1}{2} m_j^2 a_j^{-2} \int_z \Phi^2(z) +
                \frac{1}{2} \beta_j \int_z (\nabla \Phi )^2(z)
\nonu\\ &+&
                  \frac{1}{4!} \lambda_j a_j^{d-4} \int_z \Phi^4(z) +
                    \frac{1}{6!} r_j a_j^{2d-6} \int_z \Phi^6(z)
                  + \ldots
                   \enspace .
\eea
The RG flow of the running coupling constants $m_j^2$, $\beta_j$,
$\lambda_j$ and $r_j$ is approximativly given by
\bela \label{FlowEq}
m_{j-1}^2 &=&
    L^2 m_j^2 + c_{21} \lambda_j - c_{22}\lambda_j^2 + O(\lambda_j^3)
\nonu\\
\beta_{j-1} &=& \beta_j -c_{2^{\prime}2} \lambda_j^2 + O(\lambda_j^3)
\nonu\\
\lambda_{j-1} &=& L^{4-d}\lambda_j - c_{42}\lambda_j^2 +O(\lambda_j^3)
\nonu\\
r_{j-1} &=& L^{6-2d}r_j -c_{62} \lambda_j^2 + O(\lambda_j^3).
\eela
$c_{mk},$ $m\in \{ 2,2^{\prime},4,6\} ,$ $k\in \{ 1,2\} $ are constants
which depend on $j$ and $n$ ($c_{21}$ depends only on $j$). For small
$\lambda_j$, we see that $m_j^2$ is growing after RG steps. We call such
a coupling constant {\em relevant}. The coupling constant $\beta_j$
does not change (if we neglect $O(\lambda_j)$-terms) after RG steps.
Such a coupling constant is called {\em marginal}. For $d>3$ dimensions the
coupling constant $r_j$ becomes smaller after a RG step. Such a coupling
constant is called {\em irrelevant}. We see by eq.~(\ref{FlowEq}) that the
coupling constant $\lambda_j$ is relevant, marginal and irrelevant
for $d<4$, $d=4$ and $d>4$ dimensions respectively.

Let us discuss the RG flow eqs.~(\ref{FlowEq}). Consider the case
$2\le d \le 4.$ Let us choose
\be
\lambda_n = L^{(d-4)n} \lambda
\ee
for the bare coupling constant. If we iterate the RG flow
eq.~(\ref{FlowEq}) for $m_j^2$, we obtain $m_j^2 = m_j^2(\lambda_n,m_n^2).$
Each RG step produces a factor $L^2$ for $m_j^2.$ Thus after $n-j$ RG steps
we get a factor $L^{2(n-j)}.$ To perform the ultraviolet limit
$\lim_{n \rightarrow \infty }$ one has to dominate the factor $L^{2n}.$
A term of order $\lambda_n^k$ delivers a factor $L^{(d-4)kn}$ and therefore
dominates $L^{2n}$ if $(4-d)k>2.$ Thus terms of order $\lambda^k$ for
$k>\frac{2}{4-d}$ are not ``dangerous'' for the ultraviolet limit.
For example, in $d=2$ dimensions only the term $c_{21}\lambda_j$ on the
right hand side of eq.~(\ref{FlowEq}) for $m_j^2$ produces divergent terms.
To avoid such divergent terms on has to subtract this term from the
starting coupling constant $m_n^2.$ Such a subtraction term is called
{\em counter term}. Generally, in $d<4$ dimensions
the counter terms can be expressed by a finite number of Feynman graphs.
We call a model (ultraviolet) {\em super-renormalizable} if such a procedure
is possible. The
$\Phi_d^4$-model is super-renormalizable in $d<4$ dimensions.
In $d=4$ there are no suppression factors $L^{-\alpha }$, $\alpha >0,$
coming from terms containing powers of $\lambda .$ Therefore, the
counter terms cannot be expressed by a finite number of Feynman-graphs.
Nevertheless, only a finite number of coupling constants are
relevant. This property is called {\em strict renormalizability} of the
$\Phi_4^4$-model. For an existence proof of the infrared limit
$\lim_{j\rightarrow -\infty}$ for the $\Phi_4^4$-model one starts with
coupling constants $m_0^2, \beta_0, \lambda_0.$ The RG flow of
the running coupling constant $\lambda_j$ implies (for $\lambda_0$ small)
\be
\lambda_j = O(\frac{1}{|j|}).
\ee
Therefore, $\lim_{j\rightarrow -\infty} \lambda_j = 0$ and the
$\Phi_4^4$-model becomes trivial in the infrared limit. Since
$\sum_{j=0}^{-\infty} \lambda_j^k <\infty $ for $k>1$, we see that
terms of order $\lambda^k$, for $k>1$, are ``harmless'' for the
RG flow of running coupling constants.
\subsection{Excursion: Hierarchical RG fixed points}
A special application of the exact beta-function technique is
the determination of hierarchical RG fixed points. The
hierarchical RG transformations (HRGT) in $d$ dimensions is given by
a mapping $\cR \cG :\, \{ Z:\bR \rightarrow \bR \} \longrightarrow
\{ Z:\bR \rightarrow \bR \} $ defined by
\be
        Z^{\prime} (\Psi ) := \cR \cG (Z)(\Psi ) := \left[ \iGaug
         Z(\Phi +L^{1-\frac{d}{2}}\Psi ) \right]^{L^d}\enspace ,
\ee
where the Gaussian measure $\Gaug $ is defined by, $\g >0,$
\be
\Gaug  := (2\pi \g )^{-1/2} d\Phi \, e^{-\frac{\Phi^2}{2\g}}\enspace .
\ee
There exists fixed points $Z^*$, $Z^* = \cR \cG (Z^*),$ such that the
corresponding interactions $V^* := -\ln Z^*$ are $l$-wells, $l\in
\{ 2,3,\ldots \} .$ It is well-known that the $l$-well fixed points
exist in $d$ dimensions, $2\le d\le d_* = \frac{2l}{l-1}$
(cp.~ Collet and Eckmann ($\epsilon$-expansion)\cite{CE77,CE78}),
Felder (RGDE) \cite{F87}).
A first rigorous  construction of the 3-dimensional 2-well fixed point was
accomplished by Koch and Wittwer \cite{KW86,KW91}.

For a representation of partition
functions choose the following coordinates, supposing $\beta \ne 1,$
\be
Z(\varphi ) = \sum_{n=0}^\infty \frac{z_n}{\gp^n}
      :\varphi^{2n}:_{\gp } \
         \leftrightarrow \  z=(z_0,z_1,\ldots )\enspace ,
\ee
where $\gp := \frac{\g}{1-\beta^2}.$ The normal (Wick) ordering is defined by
\be
:\varphi^{2n}:_\g  := \exp \{ -\frac{\g }{2}
     \frac{\partial^2}{\partial \varphi^2} \} \, \varphi^{2n}\enspace .
\ee
Define an associative $\times$-product on $\bR^\infty $ by
\be
        (a\times b)_l = \sum_{m,n:\, |m-n| \le l\le m+n}
          \cC^{mn}_l a_m b_n\enspace ,
\ee
where the structure coefficients $\cC^{mn}_l$ obey
\be
      :\varphi^{2m}:_\g \cdot :\varphi^{2n}:_\g =
       \sum_{l:\, |m-n| \le l\le m+n}
       \g^{m+n-l}
       \cC_l^{mn} :\varphi^{2l}:_\g \enspace .
\ee
Then, the HRGT can be represented as an $L^d$-fold product
(cp. \cite{PPW94,PW94})
\be
        \cR \cG (Z) = \cS_\beta (
         \underbrace{z\times \cdots \times z}_{L^d\  factors})\enspace ,
\ee
where $\beta = L^{1-\frac{d}{2}}$ and $\cS_\beta $ is defined by
\be
(\cS_\beta (z))_l := \beta^{2l} z_l\enspace .
\ee
For the special case $L^d = 2,\ \beta = L^{1-\frac{d}{2}} =
2^{-\frac{d-2}{2d}}$
the fixed points $z^*$ are solutions of the quadratic equation
\be
z=z\times_\beta z := \cS_\beta (z\times z)\enspace .
\ee
For $\beta \ne 1$ the product $\times_\beta$ is nonassociative.

Three solutions of the fixed point
equation are immediately found. They are $0,1$ and the
high-\-tem\-pe\-ra\-tu\-re fixed point $z_{HT} =
\cN \, e^{-c_*\varphi^2}$.

There is a simple argument by $\epsilon$-expansion to show that
new non-trivial fixed points appear below $d_*=\frac{2l}{l-1}$ dimensions.
Suppose that $z_* = \eins + h$, where $\eins := (1,0,0,\ldots ),$
is a fixed point, i.~e.
\be
h = 2\eins \times_\beta h + h \times_\beta h \enspace .
\ee
This is equivalent to
\ble \label{hth}
Uh = h\times h
\ele
where $U$ is a diagonal matrix, $U= diag(1-2\beta^{2l})_{l=0,1,\ldots }.$
By eq.~(\ref{hth}) $h$ can only be infinitesimal small if
approximatively $h\in \ker U.$
But the kernel of $U$ is only non-trivial, $\ker U \ne \{ 0\} $, if
$1=2\beta^{2l}$ or equivalently $d=\frac{2l}{l-1}.$ In this case one sees
that in lowest order $h_m = -\alpha \delta_{m,l},$ i.~e. the fixed point
partition function is
\be
Z(\varphi ) = 1 - \alpha :\varphi^{2l}:_\gamma + \ldots \enspace .
\ee
Below the critical dimensions $d=\frac{2l}{l-1}$ all terms
$:\varphi^{2m}:_\gamma $ for $m\le l$ become relevant.

Koch and Wittwer extracted the
high-\-tem\-pe\-ra\-tu\-re
fixed point out of the partition function. Then $\g $ and $\beta $ change
to $L^{-2}\g $ and $L^{-2}\beta $ respectively. For their proof of the
existence of the 2-well fixed point in 3 dimensions they used the norm,
for the case $2\beta^2 <1$,
\be
        \Vert z\Vert_\rho^{(1)} := \sum_{n=0}^\infty \sqrt{(2n)!} |a_n|
            \rho^n\enspace .
\ee
For $\epsilon$-expansion one has to consider the case $2\beta^2 >1$
and uses the norm
(cp.~Pordt and Wieczerkowski \cite{PW94})
\be
        \Vert z\Vert_\rho^{(\infty )} := \sup_{n}( n! |a_n|\rho^n )
          \enspace .
\ee
In constructive field theory (cp. \cite{BY90}) a large field regulator
is used
\be
        \Vert z\Vert_\rho^{(c)} := \sup_{\phi \in \bR}
         |e^{\rho \phi^2} Z(\phi )|\enspace .
\ee
These norms are algebra-norms for
special values of $\rho .$ This means that, for all $a,b,$
\be
\Vert a\times_\beta b\Vert_\rho \le
      \Vert a\Vert_\rho \cdot \Vert b\Vert_\rho \enspace .
\ee
The beta-function technique works for $N\ge N_0 = O(1) (=7)$ using the
projection operator ${\rm P}(z) := (z_0,\ldots ,z_N,0,0,\ldots ),$ for $z\in
\bR^\infty .$
%
%
\section{Relevant and Irrelevant Parts, Renormalization,
Re\-po\-ly\-me\-ri\-za\-ti\-on}
The choice of a suitable projection operator and
the definition of running coupling constants is more complicated
for the full model than for the hierarchical case.
The choice of running coupling constants
is model-dependent and we consider in this section only
the case of the $\Phi^4$-model. The way of choosing the running coupling
constants is equivalent to an implemention of renormalization conditions.
It can be shown that renormalization is only necessary for small polymers
(cp.~Brydges and Yau \cite{BY90}, Dimock and Hurd \cite{DH91,DH92,DH93}).

In this section the questions of
renormalization of polymer activities and repolymerization of the
polymer system are discussed.
Firstly, specify some subsets $RP_n$ of the polymer set $Pol(\L )$
for $n\in \{ 0,2,2^{\prime},4\} $.
The sets $RP_n$ are called {\em renormalization parts}.
Only polymers which are contained in the renormalization parts are
renormalized. The renormalization parts obey the following
conditions

\begin{enumerate}
\item $Q\in RP_n \wedge P\subseteq Q \Rightarrow P\in RP_n$.
\item $\forall y\in \L \, \exists U_y \in Pol(\L ) :
        \forall P \in RP_n, y\in P:\, P\subseteq U_y$.
\item $RP_0 \supseteq RP_2 \supseteq RP_{2^{\prime}} \supseteq RP_4
        \supseteq \cdots $.
\item $RP_n$ preserves lattice symmetry.
\end{enumerate}
For a polymer $Y\in Pol(\L ),$ choose a function $\delta V_Y(y|\Psi )$
such that $\delta V_Y =\eins_1 \cdot \delta V_Y.$ Then, define a polymer
activity $R\in \cA$ by
\ble \label{RDef}
        Z(Y|\Psi ) = EXP(\delta V_Y + R)(Y|\Psi )\enspace .
\ele
We can choose $\delta V_Y$ in such a way that $R$ obeys
the following renormalization conditions
\ble \label{Rencon}
        R(Y_0|\Psi )\vert_{\Psi =0}  = 0,\qquad \sum_{\yp :\, \yp \in Y}
          \frac{\partial^2}{\partial \Psi (y) \partial \Psi (\yp ) }
            R(Y_2|\Psi ) \vert_{\Psi =0} = 0, \ldots \enspace ,
\ele
for all $Y_n \in RP_n,$ $\yp \in Y_n.$
$R$ is called {\em irrelevant activity}.
The functions $\delta V_Y$ correspond to perturbative
counter terms. We may compute the counter terms $\delta V_Y$ such that
$R$ defined by eq.~(\ref{RDef}) obeys the renormalization conditions
(\ref{Rencon}). For the field $\psi =0$ and the renormalization
part $Y\in RP_0,$ we have
\be
        Z(Y|0) = EXP(\delta V_Y)(Y|0) = \prod_{y\in Y} \delta V_Y(y|0)
        \enspace .
\ee
This implies
\be
        \delta V_Y(y|0) = \exp \{ \sum_{P:\, y\in P \subseteq Y}
          \frac{\widetilde{\ln Z} (P|0)}{|P|} \} \enspace ,
\ee
where the Moebius transform $\widetilde{\ln Z} $ of $\ln Z$
is implicitly defined by
\ble \label{MoeDef}
        \ln Z(Y|\psi ) = \sum_{P:\, P\subseteq Y} \widetilde{\ln Z} (P|
          \psi )\enspace ,
\ele
for all $Y\in Pol(\L ).$ Eq.~(\ref{MoeDef}) defines $\widetilde{\ln Z}$
uniquely. We have $\widetilde{\ln Z}(\emptyset ) = \ln Z(\emptyset ).$
Suppose that $\widetilde{\ln Z}(P)$ is defined for all $P\in Pol(\L )$
with $|P| < N.$ We want to define $\widetilde{\ln Z}(Y)$ for $Y\in Pol(\L )$
with $|Y| = N.$ Eq.~(\ref{MoeDef}) implies
\ble \label{RecMoe}
        \widetilde{\ln Z}(Y) = \ln Z(Y)-
         \sum_{P:\, P\subset Y\atop |P|<N} \widetilde{\ln Z} (P)\enspace .
\ele
Since the terms on the right hand side of eq.~(\ref{RecMoe}) are uniquely
defined, $\widetilde{\ln Z}(Y)$ is uniquely defined.

Eqs. (\ref{RDef}) and (\ref{Rencon}) imply,
for all $Y\in RP_2\subseteq RP_0,$
\bea
               \lefteqn{\frac{1}{2} \sum_{y^{\prime} \in Y}
               \frac{\partial^2}{\partial \psi (y) \partial \psi (
               y^{\prime})} Z(Y|\psi )\vert_{\psi =0} = }
               \nonu\\ & &
               \frac{1}{2} \sum_{P:\, y\in P \subseteq Y}
               \sum_{y^{\prime} \in P}
               \frac{\partial^2}{\partial \psi (y) \partial \psi (
               y^{\prime})} A(P|\psi )\vert_{\psi =0} =
                \nonu\\  & &
               \frac{1}{2} \sum_{y^{\prime} \in Y}
               \frac{\partial^2}{\partial \psi (y) \partial \psi (
               y^{\prime})} \delta V_Y(y|\psi )\vert_{\psi =0}
               \enspace .
\eea
We have used here that $\frac{\partial}{\partial \psi (y)} A(P|\psi )
|_{\psi =0} =0,$ which follows from the symmetry property
$A(P|-\psi ) = A(P|\psi ).$
This implies, for the counter term,
\bea
              \delta V_Y (y|\psi ) &=& \delta V_Y (y|0) +
                 \frac{1}{2} \sum_{P:\, y\in P \subseteq Y}
               \nonu\\ & &
               \sum_{y^{\prime} \in P}
               \frac{\partial^2}{\partial \psi (y) \partial \psi (
               y^{\prime})} A(P|\psi )\vert_{\psi =0} \, \psi^2(y) +
               \cdots \enspace .
\eea
We want to compute the irrelvant polymer activity $R$ for a given
counter term $\delta V_Y.$
Eq.~(\ref{RDef}) implies
\ble \label{Rdef2}
        EXP (\delta V_Y) \circ EXP (R)(Y) = EXP(A)(Y)\enspace .
\ele
Let $\widetilde{\delta V}$ be the Moebius transform of $\delta V$
\ble \label{MoeCou}
        \delta V_Y = \sum_{P:\, P\subseteq Y} \widetilde{\delta V}_P
        \enspace .
\ele
Since $EXP(A+B)=EXP(A)\circ EXP(B)$ and $\circ $ is an associative
product, eq.~(\ref{Rdef2}) implies
\ble \label{Rdef3}
        EXP(R)(Y) = EXP (-\delta V_Y) \circ EXP (A)(Y)\enspace .
\ele
This implies the following explicit formula for the irrelevant polymer
activity $R$
\bela \label{Rrep}
R(Y) &=& \sum_{P:\, P\subseteq Y} \sum_{y\in P \mapsto P_y \subseteq Y \atop
   y\in P_y} \sum_{n\ge 0} \frac{1}{n!} \sum_{P_1,\ldots P_n\subseteq
 Y-P:\,  (P_1+
\cdots +P_n) \cup \bigcup_{y\in P} P_y = Y \atop \g (P_1,\ldots ,P_n,
P_y,y\in P)\ conn.}
\nonu\\ & &
\prod_{y\in P} \widetilde{\delta V}_{P_y}(y) \,
\prod_{a=1}^n A(P_a)\enspace .
\eela
The second sum on the right hand side of eq.~(\ref{Rrep}) is over all
polymers $P_y$, for all $y\in P$, such that $y\in P.$
Since renormalization concerns only small polymers,
we have for a polymer $Y$ large
enough, the relation $\delta V_Y = \delta V$, where
\bea
        \delta V(y|\Psi ) &:=& \sum_{P:\, y\in P}
            \widetilde{\delta V}_P(y|\Psi ) \nonu\\ &=&
        c_0 + c_2\Psi^2(y) + c_{2^{\prime}} (\nabla \Psi)^2(y)
         + c_4\Psi^4(y)\enspace .
\eea
$\delta V(y|\Psi )$ is called the {\em relevant part} of the
interaction.

If one wants stability bounds or norms with large field regulators
one has to replace the counter term $\delta V(y|\Psi )$ by
$\exp \{ \delta V(y|\Psi )\} .$ The philosophy of renormalization
and repolymerization stays the same as discussed in this section.

The advantage in using this repolymerization procedure is that the
relevant part depends on a fewer number of parameters. Therefore,
the beta-function is defined on a space of lower dimension than the
space of partition functions.
The counter terms $\delta V_Y$ depend on all renormalization parts which
are contained in the polymer $Y.$ The relevant part $\delta V$ depend
on fewer terms. For the example of the $\Phi^4$-model it is determined
by the four running coupling constants
$c_0, c_2, c_{2^{\prime}}, c_4.$

To control the flow of
the irrelevant activites $R$ and the running coupling constants, we
introduce a 5th RG step. This step is called {\em repolymerization step}.
It changes the polymer system only for small polymers and therefore
the thermodynamic
limit is unchanged. The repolymerization step is the replacement of
the counter terms $\delta V_Y$ on the right hand side of
eq.~(\ref{RDef}) by the relevant part $\delta V$
\be
   Z(Y|\Psi ) \rightarrow Z^{rep}(Y|\Psi ) := EXP(\delta V + R)(Y|\Psi )
   \enspace .
\ee
For $y\in \L $ let $U_y$ be the polymer in the 2nd condition for
renormalization parts.
For $P\supseteq U_y,$ we have $ Z(P|\Psi ) =Z^{rep}(P|\Psi ).$
The projection operator $P$ for the beta-function method is defined by
\be
        {\rm P}(A) = \delta V, \qquad (\eins -{\rm P}) (A) = R\enspace .
\ee
%
%
\section{Norm Estimations}
For numerically
calculations one has to truncate the infinite number of irrelevant polymer
activities $R(P),$ $P\in Pol(\L ).$ Then, one has to estimate the truncation
error. The control
of the RG flows can be done by using norm estimates. The most
important problem for estimating polymer activities is the large field
contributions or equivalently large factorials. If we would not allow
large field contributions, then ordinary perturbation would converge.
Consider, for example, a partition function for the 0-dimensional $\varphi^4$
field theory
\ble \label{Zero}
        Z(\psi ) = \int d\varphi \, e^{-\frac{m^2}{2}\varphi^2
                -\lambda (\varphi
           +\psi )^4}\enspace ,
\ele
where $\lambda $ is a non-negative constant. If we integrate over all
$\varphi \in \bR $, we see that the integral will diverge for negative
$\lambda .$ Therefore, the series expansion in powers of $\lambda $ has zero
convergence radius. If we restrict the integration over $\varphi $
to the finite interval $[-K,K]$, $K<\infty $, then the integral is also
convergent for negative values of $\lambda $ and the power series
is convergent.
The large field behaviour for the partition function $Z$ is
\ble \label{Z0dim}
        Z(\psi ) \sim e^{-c\psi^2}, \qquad \psi \rightarrow \infty
        \enspace ,
\ele
where $c = O(\lambda^{1/2}).$ Thus
\be
        Z(\psi ) = \sum_{n=0}^\infty z_n \psi^{2n}, \qquad
           z_n \sim n!^{-1},\qquad n\rightarrow \infty \enspace .
\ee
For $c,\kappa ,\rho \in \bR_+,$ define the following norms
\be
        \Vert Z\Vert_{c,\kappa } := \sup_{\varphi ,\psi \in \bR :\atop
            |\psi | \le \kappa } \{ e^{c\varphi^2}\, |Z(\varphi +
              i\psi )|\}
\ee
and
\be
        \Vert Z\Vert_\rho := \sum_{n=0}^\infty \sqrt{(2n)!} |z_n| \rho^n
        \enspace .
\ee
For $c$ and $\rho $ small enough the norms of $Z$ defined by
eq.~(\ref{Zero}) are finite, $\Vert Z\Vert_{c,\kappa }<\infty $,
$\Vert Z\Vert_\rho < \infty .$ The Taylor coefficient $z_n$ of $\psi^{2n}$
for $Z_0(\psi ) = e^{-\lambda \psi^4}$ is of order $1/\sqrt{n!}.$
Therefore, the $\Vert \cdot \Vert_\rho $-norm of $Z_0(\psi )$
is not finite, $\Vert Z_0 \Vert_\rho =\infty .$
Thus, we are only allowed to consider norms of partition functions
after at least one RG step !

Before a norm for polymer activities can be defined, some notations for
multiindices have to be introduced. Call $m\in \cM :\, \L \rightarrow \bN $
a {\em multiindex} if
\be
|m| := \sum_{x\in \L }m(x) < \infty \enspace .
\ee
The
{\em support} of the multiindex $m$ is defined by $supp\, m :=
\{ x\in \L |\, m(x) \ne 0\} $. The factorial is defined by
$m! := \prod_{y\in \L } m(y)!.$ Let $\varphi :\L \rightarrow \bR $ be
a field and $m$ a multiindex. Define a power by $\varphi^m :=
\prod_{y\in \L} \varphi (y)^{m(y)}.$ For two multiindices $m,n\in \cM $
define an order relation by $m\le n$ iff $m(y) \le n(y)$,
for all $y\in \L .$ Let $A$ be a local polymer activity and write
\be
        A(P|\Psi ) = \sum_{m\in \cM :\atop supp\, m \subseteq P} a(P|m)
         \Psi^m\enspace .
\ee
For a polymer $P$ let $T(P)$ be the set of all tree graphs with
vertex set $P$. The tree bound of $P$ is defined by
\ble \label{DefTre}
        T_\mu (P) := \sup_{\tau \in T(P)} \exp \{ -\mu a^{-1}
         \sum_{(y\yp )\in \tau } \Vert y-\yp \Vert \} \enspace ,
\ele
where $\mu >0.$ $\Vert \cdot \Vert $ is the euclidean norm
in $\bR^d.$ For $0<k_1,k_2<1$, $\mu ,\rho >0$ and define the norm of
the polymer activity $A$
\be
        \Vert A\Vert_{k_1,k_2,\mu, \rho } :=
         \sup_{y\in \L } \left\{ \sum_{P:\, y\in P\in Pol(\L )}
           \sum_{m\in \cM : \atop supp\,  m \subseteq P}
             \frac{\sqrt{m!} |a(P|m)|}{k_1k_2^{|P|-1} T_\mu (P)}
              \, \rho^{|m|} \right\} \enspace .
\ee
The definition of the norm contains two sums.
A sum over all polymers $P$ which contains an element $y$ and a sum
over all multiindices $m$ whose support is contained in $P.$
The terms in the sums contain a square root of the factorial $m$.
This represents the large field behavior. The factor $k_1<1$ yields
that terms containing a large number of polymer activities $A$
are supressed. The factor $k_2<1$ yields that $A(P)$ is small
for polymers $P$ with a large number of elements. The factor
$T_\mu (P)$ yields that $A(P)$ is small for polymers $P$ with
large extension, i.e. polymers which contain elements $x,y$
such that $\Vert x-y\Vert $ is large.

The remainder of this section is technical. It concerns the question
of how to estimate general polymer activities after integration,
coarsening and localization step. By these methods one can control the
RG flow of the irrelevant polymer activities $R.$

\subsection{\em Integration Step}
For a multiindex $m :\, \L^2 \rightarrow \bN $, let us introduce the
notations
\ble \label{Defm12}
        m_1(x) := \sum_y m(y,x) , \qquad
        m_2(y) := \sum_x m(y,x)\enspace .
\ele
In the following we will use that
\be
\iGau{\Gamma}{\phi }P(\Phi +\Psi ) =
  \exp \{ \frac{1}{2} \sum_{x,y} \Gamma (x,y)
          \frac{\partial^2}{\partial \psi (x) \partial \psi (y)} \}
            P(\Psi )\enspace ,
\ee
where $P$ is a polynom.
Series expansion of the Gaussian measure yields
\be
        \exp \{ \frac{1}{2} \sum_{x,y} \Gamma (x,y)
          \frac{\partial^2}{\partial \psi (x) \partial \psi (y)} \} =
           \sum_{m:\L^2\rightarrow \bN }
            \frac{1}{2^{|m|}} \frac{\Gamma^m}{m!} \partial_\psi^{m_1+m_2}
              \enspace .
\ee
Define structure coefficients $C_\Gamma (m,n) ,$ for $ m,n:\, \L \rightarrow
\bN $ by
\be
         \iGau{\Gamma}{\phi}(\phi +\psi )^n = \sum_{m:\, m\le n}
           C_\Gamma (m,n) \psi^m \enspace .
\ee
We have, for $n:\, \L \rightarrow \bN ,$
\bea
    \iGau{\Gamma}{\phi}(\phi +\psi )^n &=&
          \sum_{k:\L^2\rightarrow \bN }
       \frac{1}{2^{|k|}} \frac{\Gamma^k}{k!} \partial_\psi^{k_1+k_2} \psi^n
        \nonu\\ &=&
        \sum_{k:\L^2\rightarrow \bN }
       \frac{1}{2^{|k|}} \frac{\Gamma^k}{k!} \frac{n!}{(n-k_1-k_2)!}
         \psi^{n-k_1-k_2}
            \nonu\\ &=&
       \sum_m \sum_{k:\L^2\rightarrow \bN :\atop n=k_1+k_2+m}
       \frac{1}{2^{|k|}} \frac{\Gamma^k}{k!} \frac{n!}{m!}\psi^m \enspace .
\eea
This implies
\be
      C_\Gamma (m,n) =
      \sum_{k:\L^2\rightarrow \bN :\atop n=k_1+k_2+m}
       \frac{1}{2^{|k|}} \frac{\Gamma^k}{k!} \frac{n!}{m!} \enspace .
\ee
Consider the polymer activity $A(P|\psi ) =
\sum_{n:\L \rightarrow \bN \atop supp\, n \subseteq P}a(P|n)\psi^n$
and the Gaussian integral $A^{\prime} = \mu_\Gamma (A).$
Suppose that $\ap $ are coefficients of the Taylor expansion
of $\Ap .$ Then
\be
        a^{\prime}(P|m) = \sum_{n:\, m\le n} C_\Gamma (m,n) a(P|n) ,\qquad
         a^{\prime} = C_\Gamma a \enspace .
\ee
The integration step is $EXP(A^{\prime}) = \mu_\Gamma (EXP(A))$. The
effective polymer activity are computed by an integration step
$A^{\prime} = \widetilde{\mu_\Gamma} (A).$
Thus
\be
         a^{\prime}(Y|m) = \sum_{Y=\sum P} \sum_{n:\, m\le n}
           C_{\Gamma ,\{ P\} }(m,n) \prod_Pa(P|n) \enspace ,
\ee
where
\be
        C_{\Gamma ,\{ P\} }(m,n) :=
        \sum_{k:\L^2\rightarrow \bN :\, n=k_1+k_2+m \atop
        \g (supp\, k , \{ P\} )\  conn. }
        \frac{1}{2^{|k|}} \frac{\Gamma^k}{k!} \frac{n!}{m!} \enspace .
\ee
For estimations the trick of exponential pinning is important.
Exponential pinning is given by the following bound. Consider
$\alpha \in \bR_+,\  b:\, \L \rightarrow \bR_+,\  m:\L \rightarrow \bN $.
The multinominial theorem implies
\be
        b^m |m|!^\alpha \le \Vert b^{1/\alpha}\Vert^{\alpha m} m!^\alpha ,
        \qquad \Vert b\Vert := \sum_y b(y)
\ee
and, for $d:\ \L^2 \rightarrow \bR_+ ,\ m :\, \L^2 \rightarrow \bN ,$
\be
        \frac{d^m}{m!} \le \frac{\Vert d\Vert^{|m|}}{\sqrt{m_1! m_2!}}
          \enspace ,
\ee
where $m_1$ and $m_2$ are above defined by eq.~(\ref{Defm12}).
For the coefficients $a(P|n)$, $\rho \in \bR_+$ and a polymer $Y$
define the following norm
\be
\Vert a(Y)\Vert_\rho := \sum_{m\in \cM } \sqrt{m!}\, |a(Y|m)|\, \rho^m
\enspace .
\ee
For the mapping $b: \L^2 \rightarrow \bR_+,$ define the norm
$\Vert b\Vert := sup_y \{ \sum_x b(y,x)\} .$ By exponential pinning
one can prove the following bound
\be
      \Vert a^{\prime}(Y)\Vert_\rho \le \sum_{Y=\sum P}
        \sup_{k:\, \g (supp\, k,\{ P\} )\  conn.}
         (\frac{|\Gamma |}{b})^k \prod_P \Vert a(P) \Vert_{2
            \sqrt{\frac{\Vert b\Vert}{2}} +\rho} \enspace .
\ee
Using the definition eq.~(\ref{DefTre}) of a tree
bound, we obtain
\bea
        \lefteqn{T_\mu (P_1) \cdots T_\mu (P_n) \le }
          \nonu\\ & & \inf_{\tau \in T_n}
         \exp \left\{ \mu a^{-1} \sum_{(ab)\in \tau} dist(P_a,P_b) \right\}
          \, T_\mu (P_1 \cup \cdots \cup P_n)\enspace .
\eea
Let $q>0$, $b:\L^2 \rightarrow \bR_+$ and
      $c := \frac{|\Gamma|}{b} e^{\Vert \cdot -\cdot \Vert }:\,
              \L^2 \rightarrow \bR_+$
and suppose that
\be
        4\frac{k_1}{k_2}\Vert c \Vert
        \, \Vert A\Vert_{e^{-q}k_1, e^{-q}k_2,\mu ,\rhop } < q^2 \enspace ,
\ee
where $\rhop := 2\sqrt{\frac{\Vert b\Vert}{2}} +\rho .$ Then
\be
        \Vert \widetilde{\mu_\Gamma}(A)\Vert_{k_1,k_2,\mu ,\rho } \le
          \frac{\Vert A\Vert_{e^{-q}k_1, e^{-q}k_2,\mu ,\rhop }}{1-
           4q^{-2}\frac{k_1}{k_2}\Vert c \Vert \Vert A
            \Vert_{e^{-q}k_1, e^{-q}k_2,\mu ,\rhop }}
\ee
and
\bea
        \lefteqn{\Vert \eins_{>1}\cdot (\widetilde{\mu_\Gamma}(A)
          -A)\Vert_{k_1,k_2,\mu ,\rho } \le } \nonu\\ & &
          \frac{4q^{-2}\frac{k_1}{k_2}\Vert c \Vert \Vert A
            \Vert_{e^{-q}k_1, e^{-q}k_2,\mu ,\rhop }}{1-
           4q^{-2}\frac{k_1}{k_2}\Vert c \Vert \Vert A
            \Vert_{e^{-q}k_1, e^{-q}k_2,\mu ,\rhop }}\,
         \Vert A\Vert_{e^{-q}k_1, e^{-q}k_2,\mu ,\rhop } \enspace .
\eea
After an integration step the constants $k_1$ and $k_2$ become larger
by a factor $e^q.$ Thus, for a polymer $P,$ we loose a factor $e^{q|P|}.$
This factor can become very large if $|P|$ is large. In the next subsection,
we will obtain a small factor for each element in the polymer
after the coarsening step in the case where polymers
are not small. The constant $\rho $ grows to $\rhop .$ For $d>2$ dimensions
this grow of $\rho $ can be dominated by the scaling factor
$L^{1-\frac{d}{2}}$ after the rescaling step.
For $d=2$ one uses that the coupling constant
$\lambda $ (for $\Phi^4$-model) grows by a factor $L^2$ after each RG step.
This gives a supression factor $L^{-\frac{1}{2}}$ for the constant $\rho $
which dominates the growing of $\rho $ to $\rhop .$
\subsection{\em Coarsening Step}
Norm estimation of the coarsening step requires to distiguish
between small and
large polymers. We call a polymer $X$ of the lattice $\Lp $ with lattice
spacing $La$ {\em small} if, for all $y,x\in X,$ the condition
$\frac{|x^\mu -y^\mu |}{La} \in \{ 0,1\} $, $\mu \in \{ 1, \ldots ,d
\} ,$ holds. Let $P_1,\ldots ,P_n$
be polymers of $\L $ such that $X:=[P_1+\cdots +P_n]$
is not small and the Venn-diagram $\gamma ([P_1],\ldots ,[P_n])$
is connected. Then,
there exists a positive $\epsilon = O(1)$ such that the following
tree estimation holds
\ble \label{TreEst}
        k_1\, k_2^{|P_i|-1}\prod_{i=1}^n T_\mu (P_i) \le
          k_1\, k_2^{|X|-1} T_\mu (X) \, (\frac{k_1}{k_2})^{n-1}\,
          (k_2+e^{-\mu L})^{\epsilon |P_1+\cdots +P_n|} \enspace .
\ele
Let $A$ be a polymer activity in $\L .$ The coarsened polymer activity
$\widetilde{[A]}$ is given by
\ble \label{Coa}
        \widetilde{[A]}(X)  = \sum_{n:\, n\ge 1} \frac{1}{n!}
          \sum_{P_1,\ldots ,P_n \in Pol(\L ): \atop
           P_1+\cdots +P_n =\oX ,\  \gamma ([P_1],\ldots ,[P_n])\
            conn.} \prod_{i=1}^n A(P_i) \enspace .
\ele
For small $k_2$ and large $L$ the factor $(k_2+e^{-\mu L})^{\epsilon }$
is small. The tree estimation eq.~(\ref{TreEst}) implies that for not
small polymers $X\in Pol(\Lp )$ the coarsened polymer activity
$\widetilde{[A]}(X)$ is suppressed by $\gamma^{|X|}$ where $\gamma $
is a small number. This small number can be used to supress the extra factor
$e^q$ for $k_1$ and $k_2$ after an integration step.

We estimate firstly the contributions to the norm of the coarsened polymer
$\widetilde{[A]}$ coming from not small polymers. Then, we estimate the
contributions coming from small polymers and the part of the right
hand side of eq.~(\ref{Coa}) where $n\ge 2.$
These two estimations will be discussed in this subsection.
It remains to estimate contributions coming from small polymers
and terms which contain at most one factor $A.$
For these estimations one has to use the
renormalization conditions to obtain supression factors. This method
is well-known and not discussed here (see for example \cite{R91}).

Define a projection operator $P_{ns}$ defined on the set of all polymer
activities on $\L $ which gives zero if applied to small polymers
\be
        P_{ns} (A)(X) := \left\{ \begin{array}{r@{\quad:\quad}l}
             A(X) & \mbox{$X$ not small} \\  0 & \mbox{$X$ small.}
              \end{array} \enspace \right.
\ee
For $d_1,\ldots ,d_n\in \bN $ such that $\sum_{a=1}^n d_a =2(n-1)$
denote by $T_n(d_1,\ldots d_n) $ the tree graph with vertex set
$\{ 1,\ldots ,n\} $, $d_i$ lines emerging from vertex $i$. Then
Cayley's Theorem counts the number of tree graphs in
$T_n(d_1,\ldots d_n) $
\ble \label{Cay}
        |T_n(d_1,\ldots d_n)| = \frac{(n-2)!}{\prod_{i=1}^n(d_i-1)!}
         \enspace .
\ele
Cayley's Theorem and the estimation
\bea
         \frac{|P_{ns}(\widetilde{[A]})(X)|}{k_1k_2^{|X|-1}T_\mu (X)} &\le &
          \sum_{n:\, n\ge 1} \frac{1}{(n-1)!} \sum_{d_1,\ldots ,d_n:\atop
           \sum d_i = 2(n-1)} \sum_{\tau \in T_n(d_1,\ldots d_n)}
            \nonu\\ & &
           \sum_{P_1,\ldots ,P_n :\, x\in [P_1] \atop
            \g([P_1],\ldots ,[P_n]) \supseteq \tau , \
             P_1+\cdots P_n =\oX } \prod_{i=1}^n \frac{A(P_i)}{
             k_1k_2^{|X|-1}\, T_\mu (P_i)}
            \nonu\\ & &
             (\frac{k_1}{k_2})^{n-1}\,
              (k_2+e^{-\mu L})^{\epsilon |P_1+\ldots +P_n|} \enspace ,
\eea
imply the following bound. Suppose that
$4\frac{k_1}{k_2}
      L^d \Vert A \Vert_{\kappa^{-1} k_1,\kappa^{-1} k_2,\mu ,\rho } < q^2$
holds, where $\kappa := e^{-q}(k_2 +e^{-\mu L})^\epsilon $.
Then
\be
        \Vert P_{ns}(\widetilde{[A]})\Vert_{k_1,k_2,\mu ,\rho } \le
          \frac{\Vert A \Vert_{\kappa^{-1} k_1,\kappa^{-1}
                k_2,\mu ,\rho }}{1-
           4q^{-2} \frac{k_1}{k_2}
           L^d \Vert A \Vert_{\kappa^{-1} k_1,\kappa^{-1} k_2,\mu ,\rho }}
           \enspace .
\ee
Denote by $\widetilde{[A]}_{>1}(X)$ the part of the right hand side of
eq.~(\ref{Coa}) which
consists of at least two $A$-factors and by $\widetilde{[A]}_1(X)$ the
part which consists of only one $A$-term
\be
        \widetilde{[A]}_{>1}(X)  = \sum_{n:\, n\ge 2} \frac{1}{n!}
          \sum_{P_1,\ldots ,P_n \in Pol(\L ): \atop
           P_1+\cdots +P_n =\oX ,\  \gamma ([P_1],\ldots ,[P_n])\
            conn.} \prod_{i=1}^n A(P_i)
\ee
\be
    \widetilde{[A]}_1(X) :=  \sum_{P: \, [P] =X} A(P) \enspace .
\ee
Suppose that
      $4C\frac{k_1}{k_2}q^{-2}
      L^d \Vert A \Vert_{k_1,k_2,\mu ,\rho } < 1.$ Then, we have
\be
        \Vert (\eins -P_{ns})(\widetilde{[A]_{>1}})
           \Vert_{k_1,k_2,\mu ,\rho } \le
          \frac{4Cq^{-2}\frac{k_1}{k_2}
      L^d\Vert A \Vert_{k_1,k_2,\mu ,\rho }}{1-
           4Cq^{-2} \frac{k_1}{k_2}
           L^d \Vert A \Vert_{k_1,k_2,\mu ,\rho }} \,
         \Vert A \Vert_{k_1,k_2,\mu ,\rho } \enspace .
\ee

\subsection{\em Localization Step}
Let $m :\, \L \rightarrow \bN $ be a multiindex and $\cA $ be the operator
defined in the block spin RG which maps fields defined on $\L $ to
fields defined on $\Lp .$ Expanding the $m$th power of $\cA \phi $
yields
\be
        (\cA \phi )^m = \sum_{n:\, \L^{\prime} \rightarrow \bN}
         D(n,m) \phi^n \enspace ,
\ee
where
\be
        D(n,m) := \sum_{G:\, \L \times \Lp \rightarrow \bN \atop
           G_1 =m,\, G_2 =n} \frac{m!}{G!} \cA^G \enspace .
\ee
This is proven by
\bea
        (\cA \phi )^m &=& \prod_{y\in \Lp } (\cA \phi )(y)^{m(y)} =
         \prod_{y\in \Lp }
         \left(\sum_{x\in \Lp } \cA (y,x) \phi (x) \right)^{m(y)}
         \nonu\\ &=&
        \prod_{y\in \Lp } \left(\sum_{G_y :\, \Lp \rightarrow \bN \atop
           |G_y| = m(y)} \frac{m(y)!}{G_y} (\cA (y,x)\phi (x))^{G_y(x)}
         \right)
         \nonu\\ &=&
         \sum_{G:\, \L \times \Lp \rightarrow \bN \atop G_1 =m}
          \frac{m!}{G!} \cA^G \phi^{G_2}\enspace .
\eea
For all multiindices $m:\, \L \rightarrow \bN ,$ let $a(m)$ be a real number.
Define, for a multiindex $n:\, \Lp \rightarrow \bN $
\be
        a^{\prime}(n) = \sum_{m:\, \L \rightarrow \bN } D(n,m) a(m)
           \enspace .
\ee
Then, we have
\bea
         \Vert a^{\prime}\Vert_\rho &\le &
          \sum_{n:\, \Lp \rightarrow \bN} \sqrt{n!} \sum_{m:\, \L
           \rightarrow \bN} |D(n,m)|\, |a(m)| \rho^{|n|}
           \nonu\\
           &\le & \sum_{m:\, \L \rightarrow \bN} \sum_{G:\L \times \Lp
             \rightarrow \bN \atop G_1=m,\  G_2=n}
              \frac{|\cA |^G}{G!} \sqrt{m!n!} \, \sqrt{m!} |a(m)|
               \rho^{|m|}\enspace .
\eea
Define, for $b:\ \L \times \Lp \rightarrow \bR_+,$ the norm
\be
\Vert b\Vert := \max (\sup_{x\in \Lp }\sum_{y\in \L } |b(y,x)|,
\sup_{y\in \L } \sum_{x\in \Lp } |b(y,x)|)
<\infty \enspace .
\ee
Exponential pinning, for a multiindex $G:\L \times \Lp
\rightarrow \bN ,$ yields
\be
 \frac{b^G}{G!} \sqrt{m!n!} \le \Vert b \Vert^{|G|}\enspace ,
\ee
where  $G_1 = m:\, \Lp \rightarrow \bN $ and $G_2 = n:\, \L
\rightarrow \bN $ are defined by eq.~(\ref{Defm12}). Thus
\bea
        \Vert a^{\prime}\Vert_\rho &\le &
          \sum_{m:\, \L \rightarrow \bN} \sum_{G:\L \times \Lp
             \rightarrow \bN \atop G_1=m,\  G_2=n}
              (\frac{|\cA |}{b})^G \Vert b\Vert^{|G|} \, \sqrt{m!} |a(m)|
               \rho^{|m|}
         \nonu\\
           &\le &
       \sup_{G} (\frac{|\cA |}{b})^G \, \Vert a\Vert_{\Vert b\Vert \rho }
        \enspace .
\eea
Let $A=LN(Z)$ be the polymer activity of $Z.$ Define a polymer activity
$A^{\prime} $ by
\be
Z(X|\cA_X\phi ) = \sum_{X=\sum Q } \prod_Q A^{\prime}
(Q|\cA_Q \phi )\enspace .
\ee
Use the representations
\be
        A^{\prime }(X|\psi ) = \sum_{m:\, \Lp \rightarrow \bN }
          a^{\prime} (X|m)\psi^m
\ee
and
\be
        A(X|\phi ) = \sum_{n:\, \L \rightarrow \bN }
          a (X|n)\phi^n\enspace .
\ee
Then, the following relation holds
\be
        a^{\prime} (X|m) = \sum_{X=\sum Q} \sum_{n:\, \L \rightarrow \bN }
          D_{\cA ,\{ Q\} } (m,n) \prod_Q a(Q|n_Q)\enspace ,
\ee
where
\be
        D_{\cA ,\{ Q\} } (m,n) := \sum_{G:\, \L \times \Lp \rightarrow \bN
          \atop G_1=m,\, G_2=n,\  \g (supp\, G, \{ Q\} )\  conn.}
            \frac{m!}{G!} \cA^G \enspace .
\ee
The coefficients $a$ and $a^{\prime} $ obey the following norm inequalities
\be
        \Vert a^{\prime}(X)\Vert_\rho \le \sum_{X=\sum Q}
          \sup_{G:\, \g (supp\, G, \{ Q\} )\  conn.}
            (\frac{|\cA |}{b})^G \, \prod_Q \Vert a(Q)\Vert_{\Vert b\Vert
              \rho }\enspace .
\ee
Define $c:= \frac{|\cA |}{b} e^{\mu \Vert \cdot -\cdot \Vert }:
      \, \L \times \Lp \rightarrow \bR_+$ and suppose, for $q>0,$
\be
4\frac{k_1}{k_2}q^{-2}\Vert c\Vert \Vert A
\Vert_{e^{-q}k_1, e^{-q}k_2, \mu ,\Vert b\Vert\rho } < 1\enspace .
\ee
Then
\be
        \Vert A^{\prime}\Vert_{k_1,k_2,\mu ,\rho } \le
          \frac{\Vert A\Vert_{e^{-q}k_1, e^{-q}k_2,\mu ,
               \Vert b\Vert \rho }}{1-
           4q^{-2}\frac{k_1}{k_2} \Vert c
              \Vert \Vert A\Vert_{e^{-q}k_1, e^{-q}k_2,\mu ,
                 \Vert b\Vert \rho }}\enspace .
\ee
%
%
\section{Conclusion}
The input of a RG transformation starts with a given relevant part
$\delta V$
depending on a finite number of (running) coupling constants and an
irrelevant polymer activity $R.$ $\delta V$ is a polymer function
vanishing for polymers which contain more than 1 element. The
polymer activity $A$ is the sum of these two terms, $A=\delta V +R.$
A new polymer activity $\widetilde{A}$ is defined after integration,
coarsening, localization and rescaling step, $\widetilde{A} :=
\widetilde{\cR_L}\circ \widetilde{\iota_\cA}\circ \widetilde{[\ ]}\circ
\widetilde{\mu_\Gamma}(A).$ For a polymer $X$ of $\Lp $ the new irrelevant
activity $\Rp $ is defined by
\be
EXP(\widetilde{A})(X) = EXP(\delta V_X + \Rp )(X) \enspace ,
\ee
where the counter terms $\delta V_X$ are defined by the polymer
activities $\widetilde{A}$ such that $\Rp $ fulfills renormalization
conditions. The new (running) coupling constants
determine $\delta \Vp .$ The sum of the new relevant part
$\delta \Vp $ and the irrelevant polymer activity $\Rp $ gives
the new effective polymer activity $\Ap .$

The RG flow is splitted into a flow of the relevant part $\delta V$ and
irrelevant polymer activity $R$
\be
\delta V \rightarrow \delta V^{\prime} (\delta V,R),\qquad
R\rightarrow R^{\prime} (\delta V,R) \enspace .
\ee

The split of the RG flow is model-dependent. The methods for a
definition and control of the irrelevant polymer activities is
model-independent.

The control of the RG flow is solved by proving recursive bounds on
the running coupling constants and the norms of the irrelevant polymer
activities $R.$ The definition of the norms has to contain parameters
which control the size and extension of polymers and terms which control
the large field behavior.

For an explicit control of the RG flow one proceeds as follows.
We start with a bare
interaction $\delta V_n$ and an irrelevant polymer activity $R_n=0.$
For $j<n$ the effective interactions $\delta V_j^{(n)}$ and irrelevant
activities $R_j^{(n)}$ are defined by the RG flows
\be
\delta V_j^{(n)} \rightarrow \delta V_{j-1}^{(n)} (\delta V_j^{(n)},
R_j^{(n)}), \qquad
R_j^{(n)} \rightarrow R_{j-1}^{(n)} (\delta V_j^{(n)},
R_j^{(n)}) \enspace .
\ee
$\delta V_j^{(n)}$ depends on certain coupling constants $\gamma_{0,j}^{(n)},
\ldots ,\gamma_{N,j}^{(n)} \in \bR .$ For the control of the RG flow one
suposes that there exists finite intervals $I_{k,j}^{(n)} \subset \bR ,$
for $k\in \{ 0,\ldots ,N\} $, $j\le n$,
and constants $k_1,k_2,\mu ,\rho ,\epsilon_j^{(n)}$, for $k\in \{ 0,
\ldots ,N\} ,\  j\le n$ such that
\be
\gamma_{k,j}^{(n)} \in I_{k,j}^{(n)}, \qquad \Vert R_j^{(n)}\Vert_{k_1,
k_2,\mu ,\rho} < \epsilon_j^{(n)}
\ee
implies
\be
\gamma_{k,j-1}^{(n)} \in I_{k,j-1}^{(n)}, \qquad \Vert R_{j-1}^{(n)}
\Vert_{k_1,k_2,\mu ,\rho} < \epsilon_{j-1}^{(n)}.
\ee
For weakly coupled models, a guess of the intervals $I_{k,j}^{(n)}$ and
constants $\epsilon_j^{(n)}$ can be achieved in the following way. Set
$R_j^{(n)}=0$ and compute the RG step
\be
\delta V_j^{(n)} \rightarrow \delta V_{j-1}^{(n)}(\delta V_j^{(n)} ,0)
\ee
using perturbation theory. This gives an approximate relation of the
flow of the coupling constants $\g_{0,j}^{(n)}.$ For a guess of the constant
$\epsilon_j^{(n)}$ use the RG step
\be
R_j^{(n)} = 0 \rightarrow R_{j-1}^{(n)} (\delta V_j^{(n)}, R_j^{(n)}=0)
\ee
and find a bound for
$ \Vert R_{j-1}^{(n)} (\delta V_j^{(n)}, R_j^{(n)}=0)\Vert_{k_1,k_2,\mu ,
\rho } .$

Iterating the RG equations one sees that the relevant interaction terms
$\delta V_j^{(n)}$ and the irrelevant activities $R_j^{(n)}$ depend on the
starting (bare) interaction $\delta V_n.$ For the infrared limit one has
to show that $\delta V_n$ can be defined such that the limits
$\lim_{j\rightarrow -\infty} \delta V_j^{(n)}(\delta V_n)$ and
$\lim_{j\rightarrow -\infty} R_j^{(n)}(\delta V_n)$ exist. For the
ultraviolet limit one has to show that the limits
$\lim_{n\rightarrow \infty} \delta V_j^{(n)}(\delta V_n)$
and $\lim_{n\rightarrow \infty} R_j^{(n)}(\delta V_n)$ exist.

\end{document}